\documentclass[eprint]{revtex4-1}

\usepackage{amsmath}
\usepackage{color}
\usepackage{chngcntr}
\counterwithin{equation}{section}
\usepackage{graphicx}
\usepackage{bbm}
\usepackage{dsfont}
\usepackage{blindtext}
\usepackage{float}
\usepackage{caption}
\usepackage{subcaption}
\DeclareGraphicsExtensions{.png,.pdf}
\usepackage{amssymb}
\begin{document}


\title{Distinguishing Thermal Fluctuations from Instrumental Error for High Pressure Charged Gas}


\author{Alek Bedroya}
\email{bedroya\underline{ }alek@physics.sharif.edu}
\affiliation{Department of Physics, Sharif University of Technology, Azadi St., Tehran, Iran}
\affiliation{Department of Mathematics, Sharif University of Technology, Azadi St., Tehran, Iran}

\author{Mahmud Bahmanabadi}
\email{bahmanabadi@sharif.edu}
\affiliation{Department of Physics, Sharif University of Technology, Azadi St., Tehran, Iran}


\date{\today}

\begin{abstract}
Thermodynamic parameters such as temperature and pressure can be defined from the statistical behavior of a system. Therefore, thermal fluctuation is an inseparable characteristic of these parameters which eventually finds its way into experimental data. Analyzing these fluctuations is very useful in studying the phase transitions of a physical system or its behavior around critical points. However, this approach is not straightforward as most of the times it is impossible to distinguish meaningful thermal fluctuations from those due to the instrumental errors. In this article, we have offered a method by which an experimenter can separate this multi-sourced fluctuation into its constitutive parts according to their sources. Although the article is only focused on a specific system, which is a high pressure charged gas, we have used a computational method which could be used for various other systems. Our proposed idea is very efficient and reduces the required computation time by a remarkable fraction. We have used Euler algorithm, which generally does not hold the internal energy conserved; But we have used this fact as a tool which allows us to surf in the phase space of the system and reach different energy levels in significantly less time. Although system does not spend enough time in a single energy level to equilibrate, but we have been able to extract the details of the equilibrium state out of our data. Using numerical computations combined with theoretical modelings we have given a final expression for the amount of the overall fluctuations existing in the measured pressure values. This expression is given in terms of the characteristics of both the gas and the barometer so that it can be experimentally verified.
\end{abstract}


\maketitle

\section{Introduction}
Almost any physical experiment includes statistical uncertainties, which make fluctuation an inseparable part of physical measurements. The amount of this uncertainty sometimes becomes a valuable source of information e.g. phase transitions. Besides statistical uncertainties, instrumental errors is another permanent source of fluctuation which has nothing to do with the statistical behavior of the system. This variation can be reduced by improving the instruments but it never reaches zero. The fluctuation of any experimental measurand includes both the statistical uncertainties and instrumental error. At this point, an important question comes to mind:

\bf Is it possible to distinguish the thermal fluctuation from the instrumental error?\normalfont

Even if not possible in general, we can propose models based on examining our instruments and systems to predict the overall fluctuations. So that by comparing these predictions with experimental results,we can verify whether we have understood the  correctly. If the proposed theory matches with experiment, then based on our model we can see its estimate for the amount of thermal fluctuation.

Following this idea, we simulate a charged gas under strong electrostatic interactions. Simulating charged gas using particle simulation methods has been done before \cite{1,2}. But as explained in section II, we are proposing a novel method which considerably accelerates the simulation.

Our setup is a two dimensional box, containing 200 charged particles with electrostatic interactions between them. The simulation is described in details in section II. We derive the dynamics of the system using Euler algorithm.

Next we model a barometer as a damped harmonic oscillator in section III where we mathematically formulate the expected properties of an ideal barometer to get a reasonable minimal theoretical model for it. Afterwards, we use this model of barometer in our computer simulation in order to obtain the instrumental fluctuation.

In section IV, we derive a theoretical expression for the thermal fluctuations based on Thermodynamics.

Finally, in section V, we combine the results for thermal and instrumental fluctuations which have been obtained in previous sections and we present a general formula for the overall fluctuation in the experimental measured values.

Please note that we rescaled our quantities by dividing them to a factor with the same dimension in order to make the plots looking better. The reason we are doing this is because we are only interested in the form of functionalities not the exact values. These constant factors are being showed by a 0 subscript through the whole article e.g. we de-dimensionalize $p$ by $p_0$.

\section{Simulation Configurations}
We assumed a two dimensional square box containing 200 identical charged particles going under electrostatic interaction and
elastic collisions with the walls. At the initial configuration, the particles had been uniformly distributed along two line segments with equal lengths and a mutual endpoint in the box and the other two endpoints being placed on two opposite corners of the box (figure 1a). The initial condition is symmetric with respect to the diagonal, but we break the symmetry by adding a small perturbation. After a short time they spread through the box and reach equilibrium (figure 1f). Note that we used Euler algorithm for this simulation. In general, Euler algorithm does not conserve energy in many body simulations similar to this. However, this would be an advantage if we could define thermal parameters during small time interval in which the system's energy does not change significantly. This way, one can find the functionality of pressure in terms of internal energy just in one run, which is very efficient in time.

\begin{figure}[H]

\begin{subfigure}{0.33\textwidth}
\includegraphics[width=.5\linewidth]{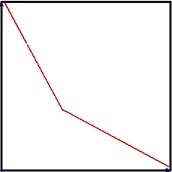}
\caption{\label{fig:sub1}}
\end{subfigure}%
\begin{subfigure}{.33\textwidth}
\includegraphics[width=.5\linewidth]{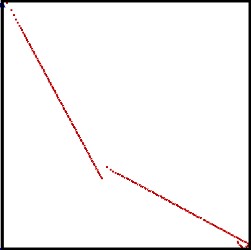}
\caption{\label{fig:sub2}}
\end{subfigure}
\begin{subfigure}{.33\textwidth}
\includegraphics[width=.5\linewidth]{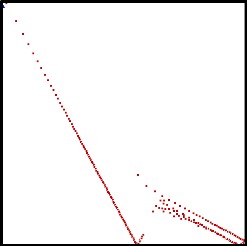}
\caption{\label{fig:sub3}}
\end{subfigure}
\begin{subfigure}{.33\textwidth}
\includegraphics[width=.5\linewidth]{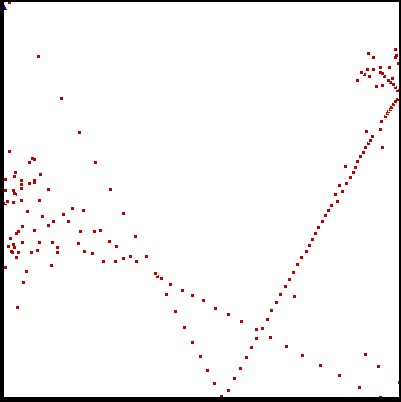}
\caption{\label{fig:sub4}}
\end{subfigure}
\begin{subfigure}{.33\textwidth}
\includegraphics[width=.5\linewidth]{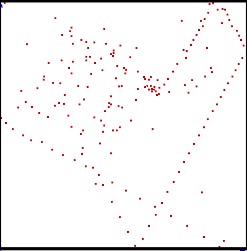}
\caption{\label{fig:sub5}}
\end{subfigure}%
\begin{subfigure}{.33\textwidth}
\includegraphics[width=.5\linewidth]{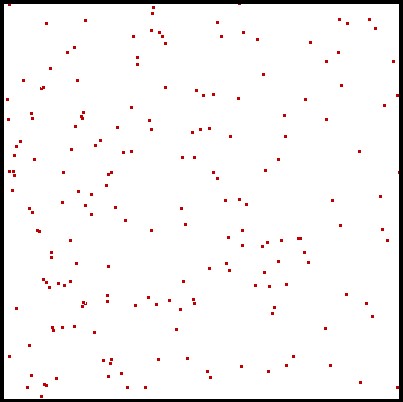}
\caption{\label{fig:sub6}}
\end{subfigure}
\caption{\label{fig:1}The evolution of the system from an initial arbitrary state to equilibrium ( a to f).}

\end{figure}

Using Euler algorithm, we can obtain the state of system in time $t$, based on its previous state in time $t-\delta t$. In order to find the horizontal (vertical) pressure, we took average over the momentum transferred perpendicularly to the horizontal (vertical) walls during the time interval $[t,t + \Delta t]$. The ratio $\Delta t/\delta t$ is a dimensionless quantity which shows the ratio of macroscopic time scale over the microscopic time scale which has set equal to 1000 in our simulation. As we mentioned, the internal energy will increase in time because of low precision of the Euler algorithm. To see whether the system spends enough time in a single energy state to reach its equilibrium, we have plotted the $p_x,p_y$ (respectively the horizontal and vertical pressures) against each other (figure 2).

\begin{figure}[H]
\centering
\includegraphics[width=0.6\linewidth]{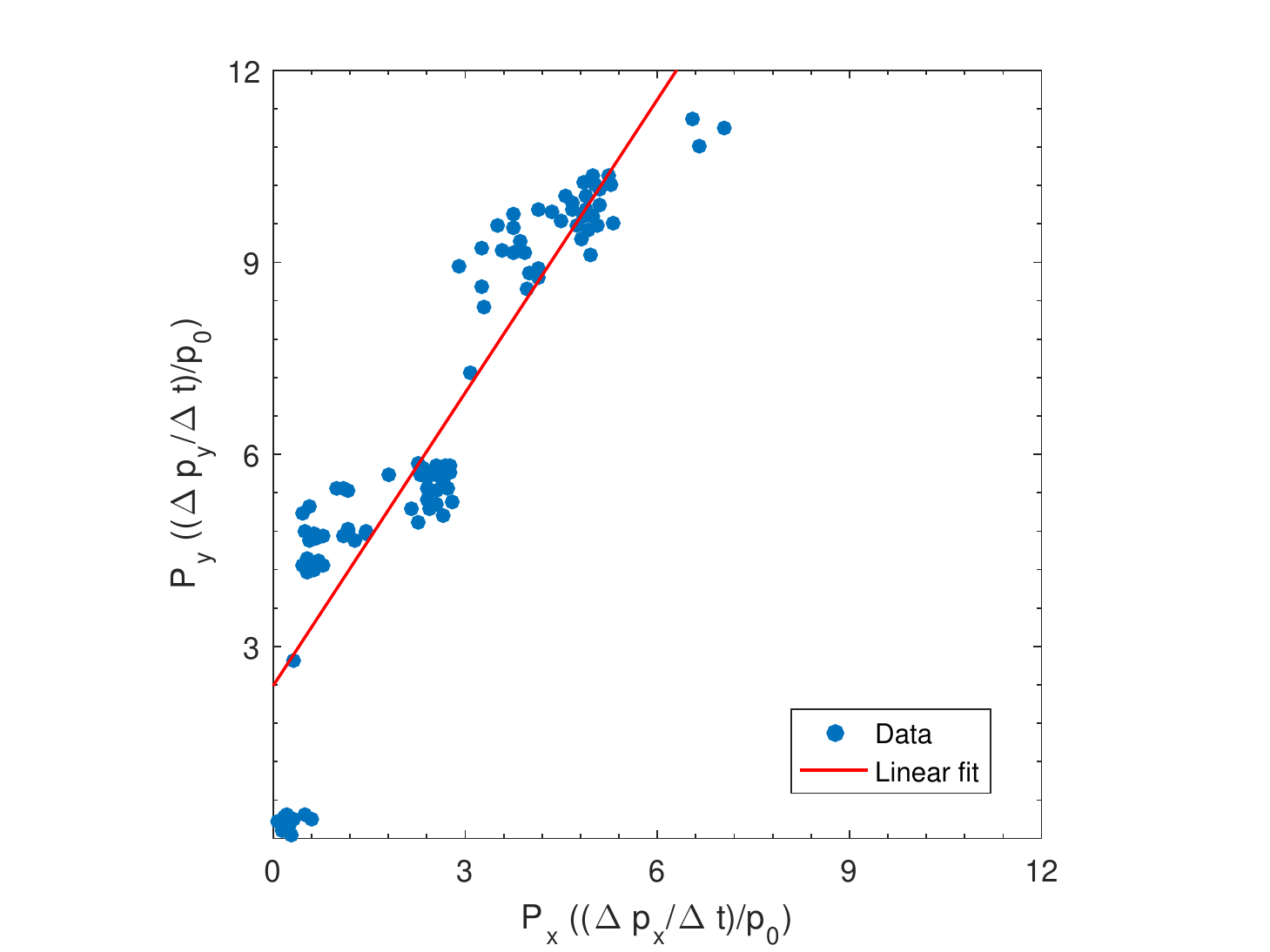}
\caption{\label{fig:2}Vertical pressure versus horizontal pressure with same units ($p_0$). Best linear fit: $p_y\approx1.53p_x+2.38p_0$, $R^2\approx0.84$}

\end{figure}
In the equilibrium state, horizontal and vertical pressures should be the same. According to figure 2, these two pressures have a linear relationship with a slope different from $1$ . This clearly indicates that the system had not reached equilibrium. Besides, the linear fit to the data of figure 2 has a non-zero intercept, which means when the pressure is zero in one direction, there is a non-zero net pressure in the other direction! In order to see whether or not two pressures can have a linear relation with zero intercept, we  rescaled and normalized each data point of figure 2 and projected it on the unit circle around origin (figure 3). Given a linear relation with zero intercept, the normalized points would fall around a single point; However as one can see in figure 3, this is clearly not the case.
\begin{figure} 
\centering
\includegraphics[width=0.6\linewidth]{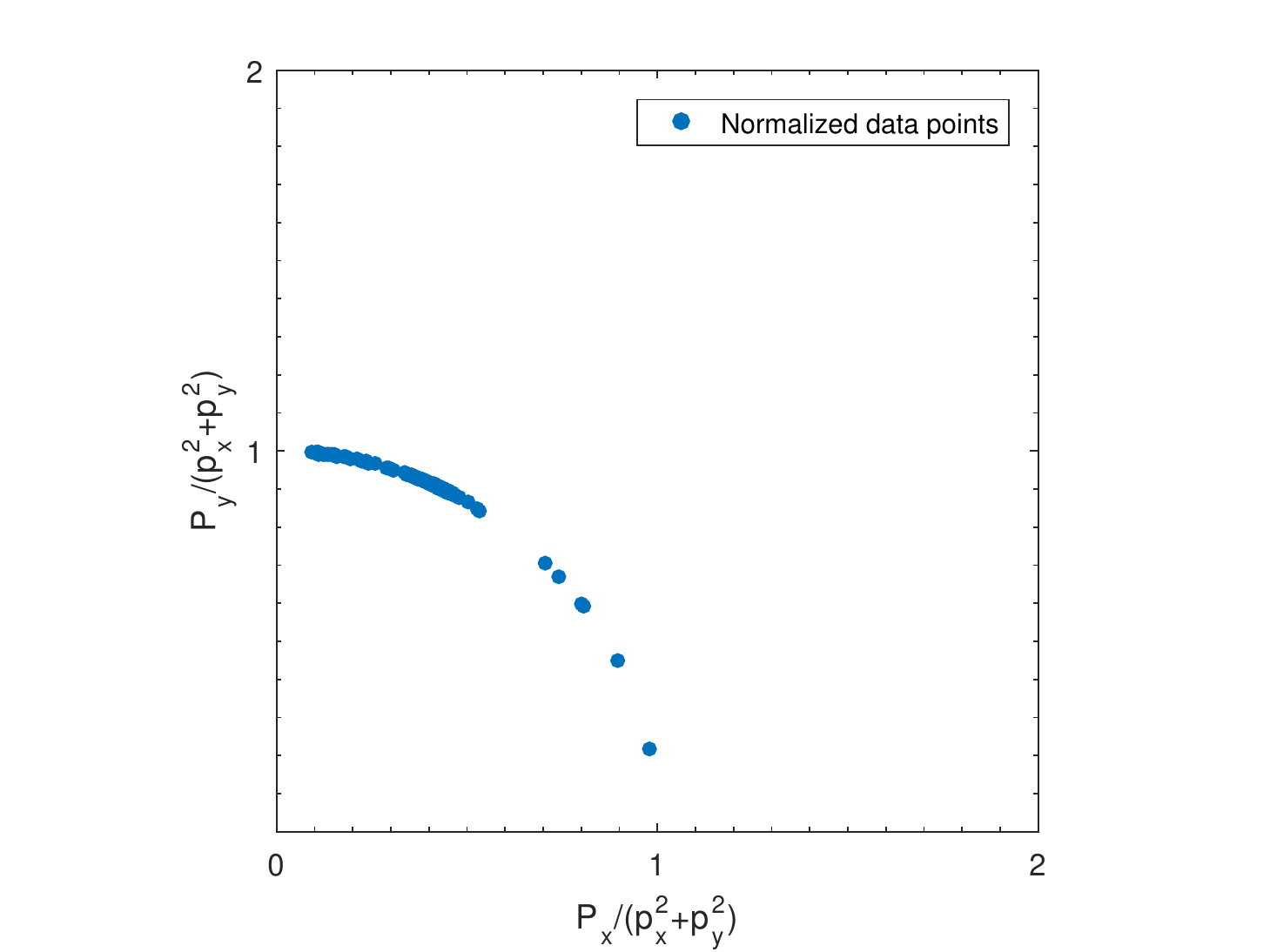}
\caption{\label{fig:3}The same figure 2 after normalizing data points}
\end{figure}
This means that because of the changes in the internal energy, system did not have enough time to reach its equilibrium. Before giving up, we make another attempt to extract the information of the equilibrium state. We consider $p=(p_x+p_y)/2$ as the first linear approximation of the equilibrium pressure and we plot the internal energy, U, in terms of p in figure 4.

Surprisingly, this time not only $p$ becomes a function in terms of $U$, in contrast to the previous case which horizontal and vertical pressures each had several values for a single energy, but also it is perfectly linear in terms of $U$ with $R$ squared of 0.98 (figure 4). These two facts support this approximation. Therefore we found for any given volume $V$, $p$ is linear in terms of $U$. One may summarize this as following,

\begin{equation}\label{eq:pyth}
U=pf(V)
\end{equation}

\begin{figure}
\centering
\includegraphics[width=0.6\linewidth]{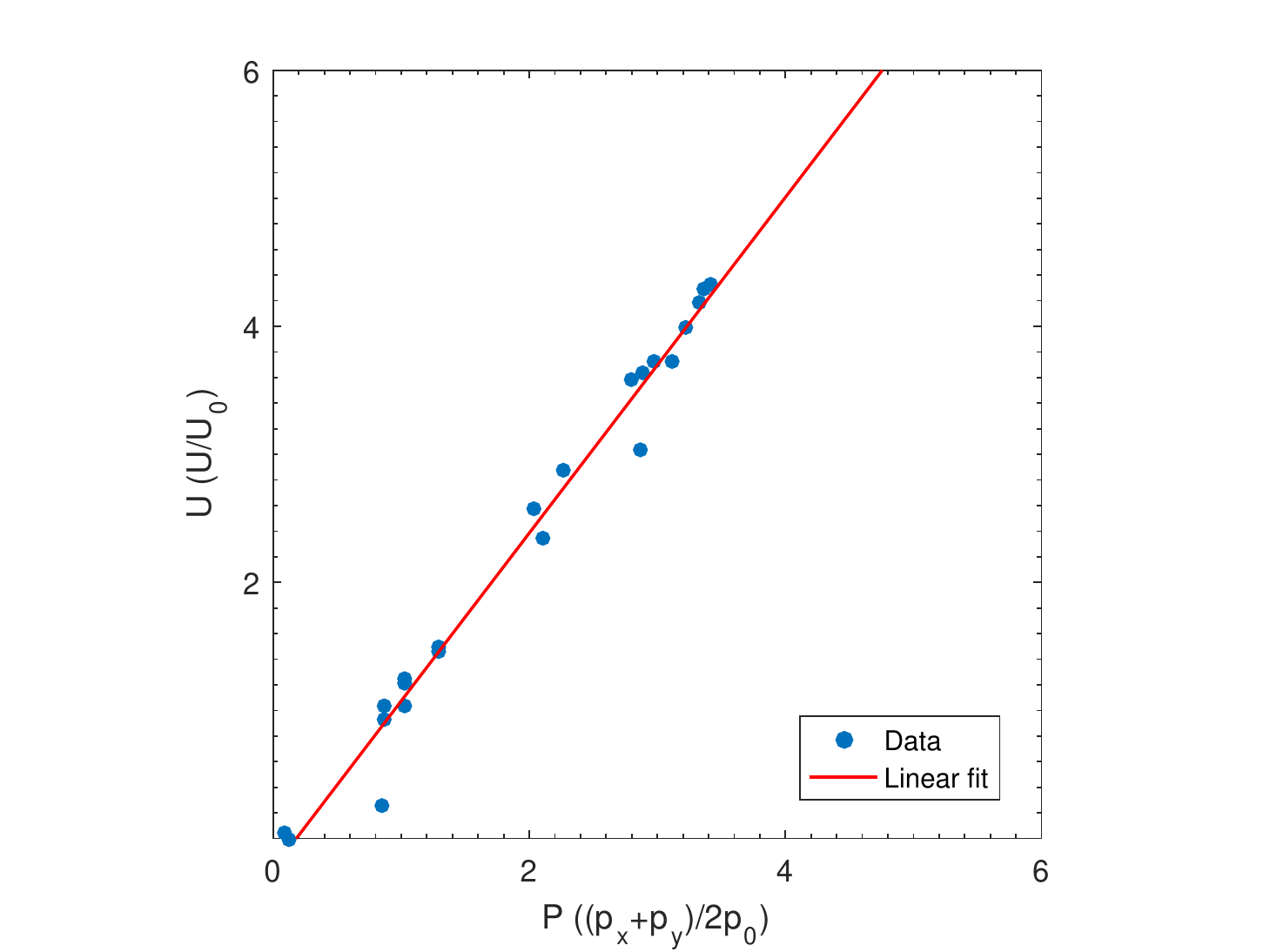}
\caption{\label{fig:4}Internal Energy, $U$, in terms of pressure, $\frac{p_x+p_y}{2}$. Best linear fit: $\frac{U}{U_0}\approx1.32\frac{p}{p_0}-0.24$, $R^2\approx0.98$}

\end{figure}
In order to obtain the function $f(V)=U/p$, we run the simulation for different volumes, $V$, and calculate the slope of $U$ in terms of $p$. Figure 5 shows the result.

\begin{figure}
\centering
\includegraphics[width=0.6\linewidth]{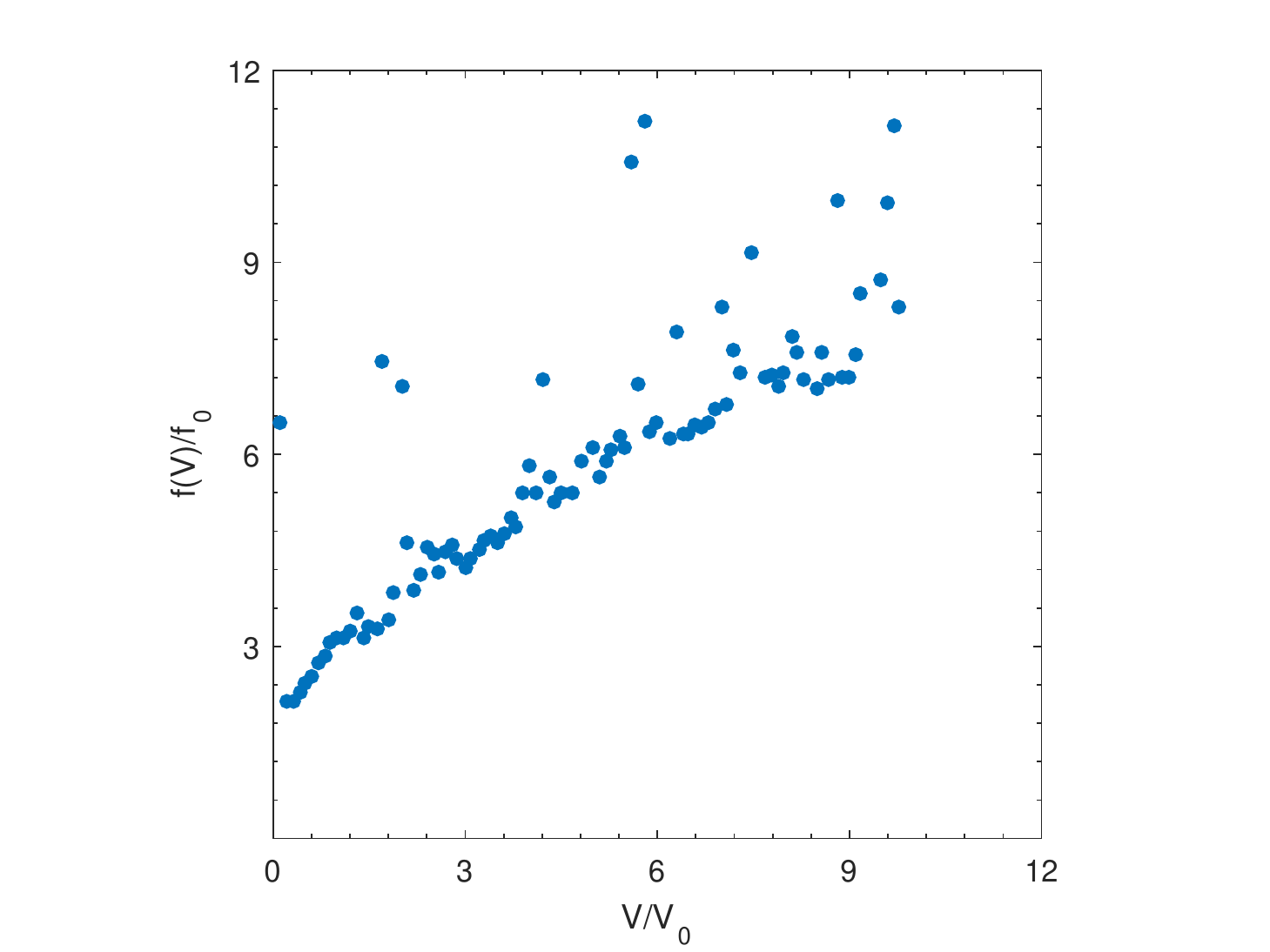}
\caption{\label{fig:5}The function $f(V)=\frac{U}{p}$ in terms of volume, $V$.}

\end{figure}

As you can see in figure 5, there are some outliers which appear as a result of using Euler algorithm. In order to delete these wrong data points, we put a discontinuity cutoff (figure 6).

As one can see in figure 6, $f(V)$ is a linear function in terms of $V$ ($R$ squared equals to $0.96$) with nonzero y-intercept. Hence one may write $U$ in terms of $p$ and $V$ as $U=p(aV-b)$. In appendix \ref{app:1}, under few mentioned assumptions we have shown that for this kind of state equation the temperature becomes proportional to the internal energy, and the equation of state has the following form:

\begin{align}
U&=p(aV-b);\nonumber\\
T&=kU\label{stateeq}
\end{align}
where $k$ is a constant.
\begin{figure}[H]
\centering
\includegraphics[width=0.6\linewidth]{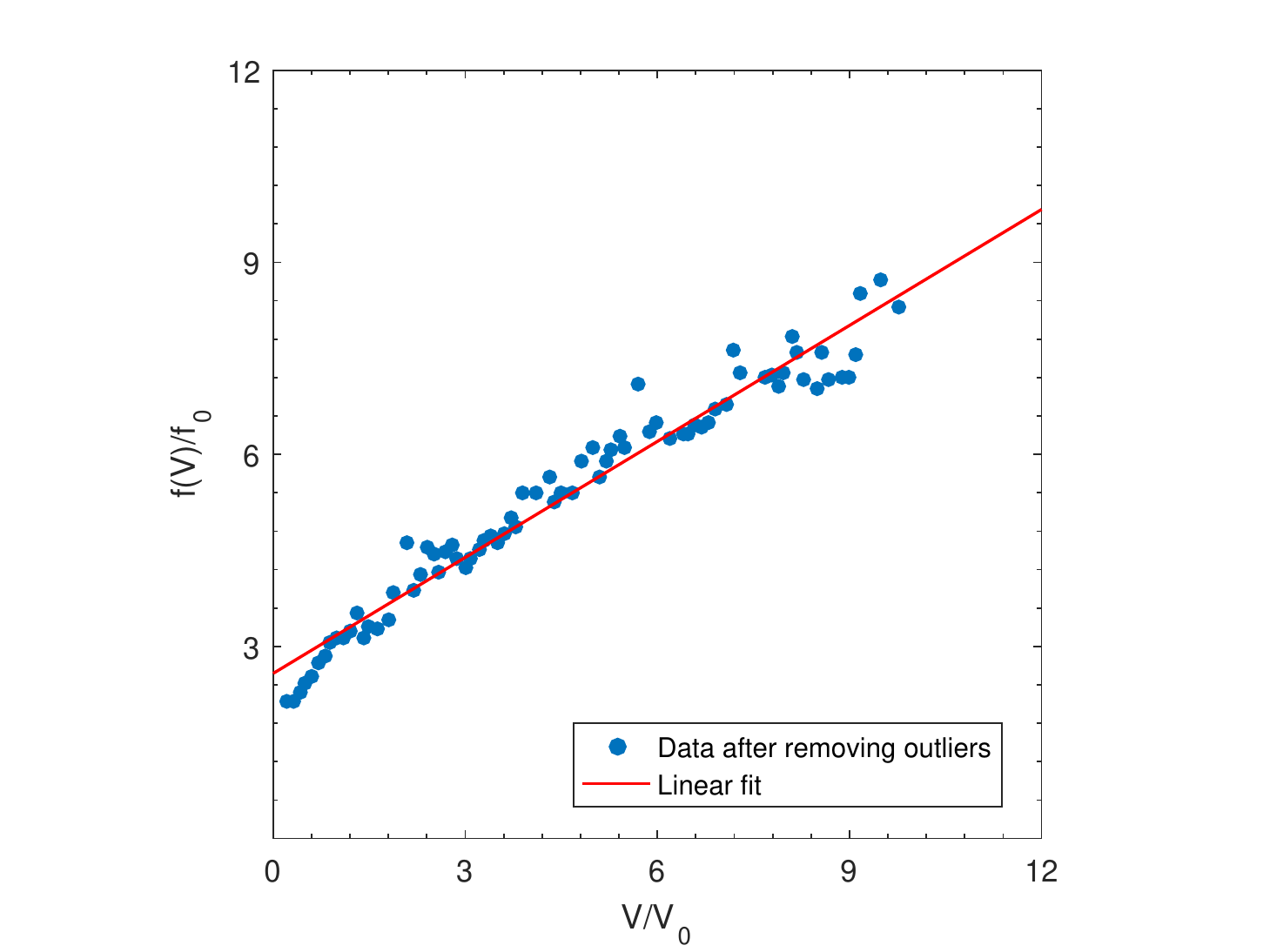}
\caption{\label{fig:6}The same figure 5 after deleting outliers. Best linear fit: $\frac{f(V)}{f_0}\approx0.60\frac{V}{V_0}+2.58$, $R^2\approx0.96$}

\end{figure}

\section{Instrumental Fluctuations}

In this section, we model a barometer by a damped harmonic oscillator. Consider that the motion of the pressure measuring surface, could be described by the following equation:
\begin{equation}
\ddot x+2\beta \dot x +\omega^2 x=0\label{oscillator}
\end{equation}
where $\beta$ is the damping constant and $\omega$ is the angular frequency of the oscillator. The case when $\beta^2<\omega^2$ is referred to as under damping oscillation. In this case, when starting with an initial condition $x =0$ and $\dot x>0$, it oscillates along time and the frequently changes its sign (figure 7) \cite{3}.

\begin{figure}[H]
\centering
\includegraphics[width=0.6\linewidth]{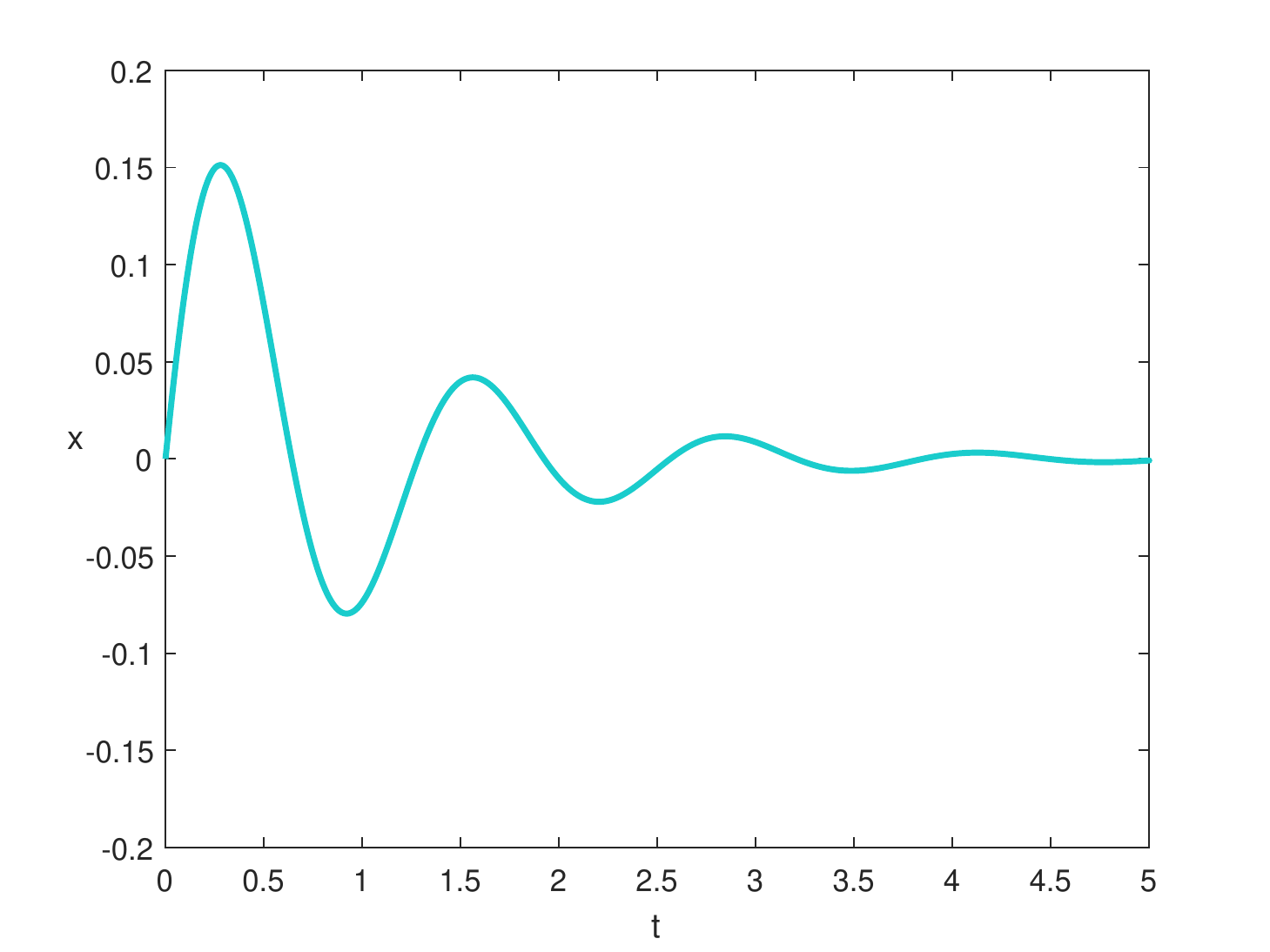}
\caption{\label{fig:7}Under damping solution ($\beta=1,\omega=5$) of the equation \ref{oscillator} with initial conditions $x =0, \dot x=1$.}

\end{figure}

The variable $x$, corresponds to the pressure of the gas which has been measured by barometer. Obviously, we do not expect a negative pressure reading from a barometer; and in order to accommodate that, it would be reasonable to assume $\beta\geq \omega$.

Every collision between a gas particle and the barometer's surface, can be considered as a discontinuous jump in the momentum of the barometer's measuring surface. Hence in order to obtain the movement of the barometer's surface, we have to solve equation \ref{oscillator} with the initial conditions $x=0,\dot x =2p/M$ where $p$ is the amount of momentum being transferred to the barometer's surface because of the collision and $M$ is the mass of the surface. Thanks to the linearity of differential equation \ref{oscillator}, it is sufficient to solve it with the initial conditions $x=0,\dot x =1$ and then multiply the solution by $2p/M$. The graph of the solution, using the mentioned values for $\beta$ and $\omega$, is plotted in figure 8.

\begin{figure}[H]
\centering
\includegraphics[width=0.6\linewidth]{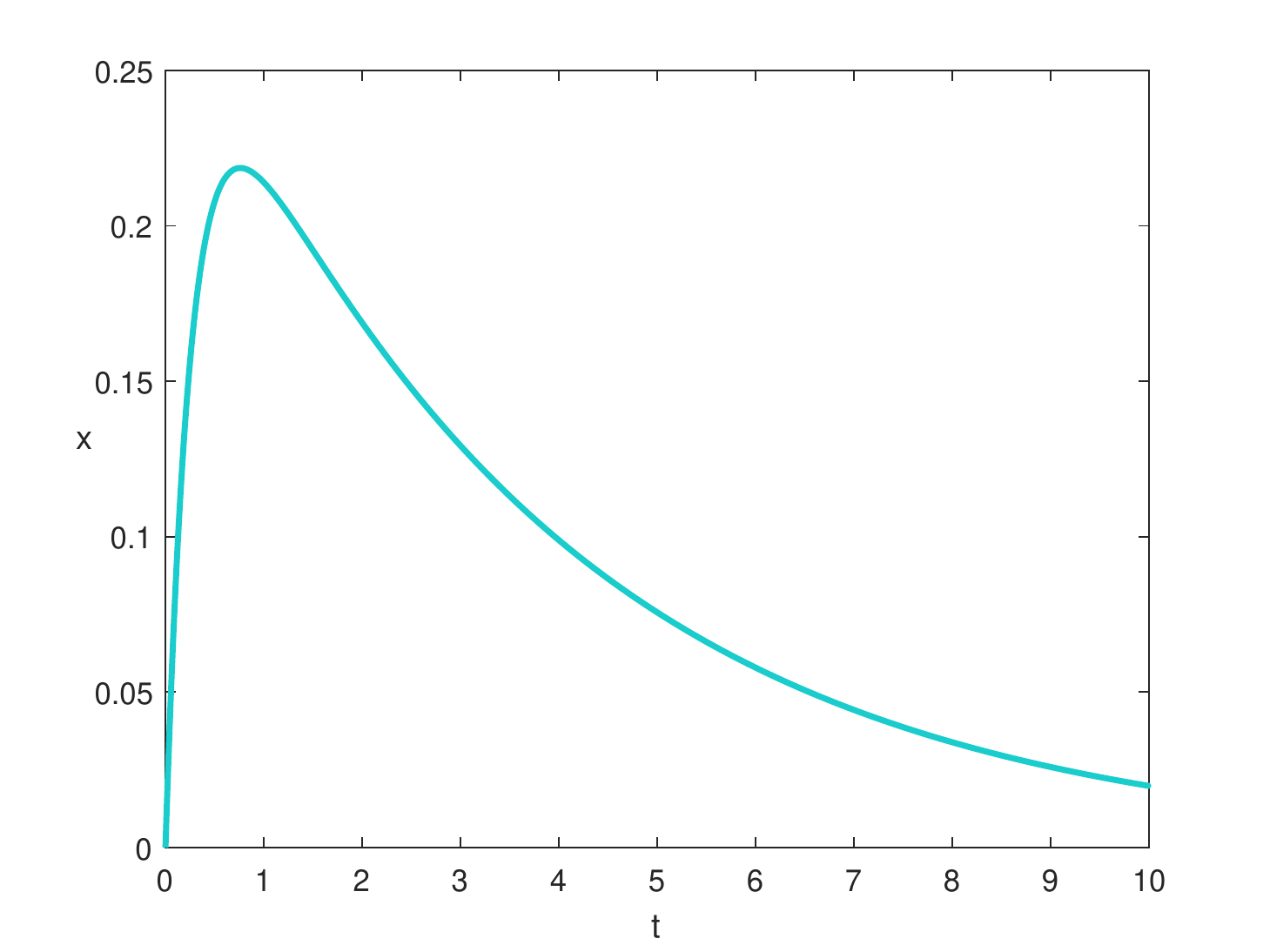}
\caption{\label{fig:8}Over damping solution ($\beta=2,\omega=1$) of the equation \ref{oscillator} with initial conditions $x =0, \dot x=1$.}

\end{figure}
Previously, we averaged the momentums transferred to the barometer's measuring surface in different times, with respect to a uniform normal weight function $f(t)$ which is equal to 1 inside the interval $[0,\Delta t]$ and vanishes outside. In fact this function is the impact of a collision to the amount of measured pressure at $t$ seconds later. Somehow $f(t)$ describes how the barometer remembers a collision in the past. Hence in order to find the pressure reported by our modeled barometer, we have to find the average transferred momentums with respect to the new weight function. In particular, we should find the normalized solution of equation \ref{oscillator} with the initial condition $x=0$. This function shows the effect of momentum transition after time $t$ in the position of barometer's surface. As shown in appendix \ref{app:2}, this unique normalized solution is given as,
\begin{equation}
x(t)=\frac{\omega^2}{2\sqrt{\beta^2-ω^2 }}(e^{(-\beta+\sqrt{\beta^2-\omega^2 })t}-e^{(-\beta-\sqrt{\beta^2-\omega^2 })t})
\end{equation}
So we ran the simulation again and recalculated the variance of measured pressure data due to the error of this modeled barometer for different values $\beta/\omega$ in the interval $[1,5]$ (figure 9). As you can see in figure 9, the instrumental fluctuation can be very well described by a quadratic polynomial ($R$-squared equals to $0.999$). However we know that this second degree polynomial approaches to $-\infty$ as $\beta/\omega$ goes to infinity. Hence it certainly cannot be considered as a good approximation for the instrumental fluctuation which is always a positive number. But for a wide range of $\beta/\omega$, including the real experimental range ($<3$), this approximation is valid. So this approximation is reasonable for an ordinary mechanical barometer.

Therefore the probability of having error $\Delta p$ in a measured pressure can be calculated as follows,
\begin{align}
Pr_{Inst.} (\Delta p)&=m' e^{-n'_\omega(\beta) \Delta p^2 }\label{instrumentaldist}
\end{align}
where $m'$ is a constant and $n'_\omega(\beta)$ is inverse of a quadratic polynomial given as,
\begin{align}
n'_\omega(\beta)&=(c_2 (\omega) \beta^2+c_1 (\omega)\beta+c_0 (\omega))^{-2}
\end{align}
\begin{figure}[H]
\centering
\includegraphics[width=0.6\linewidth]{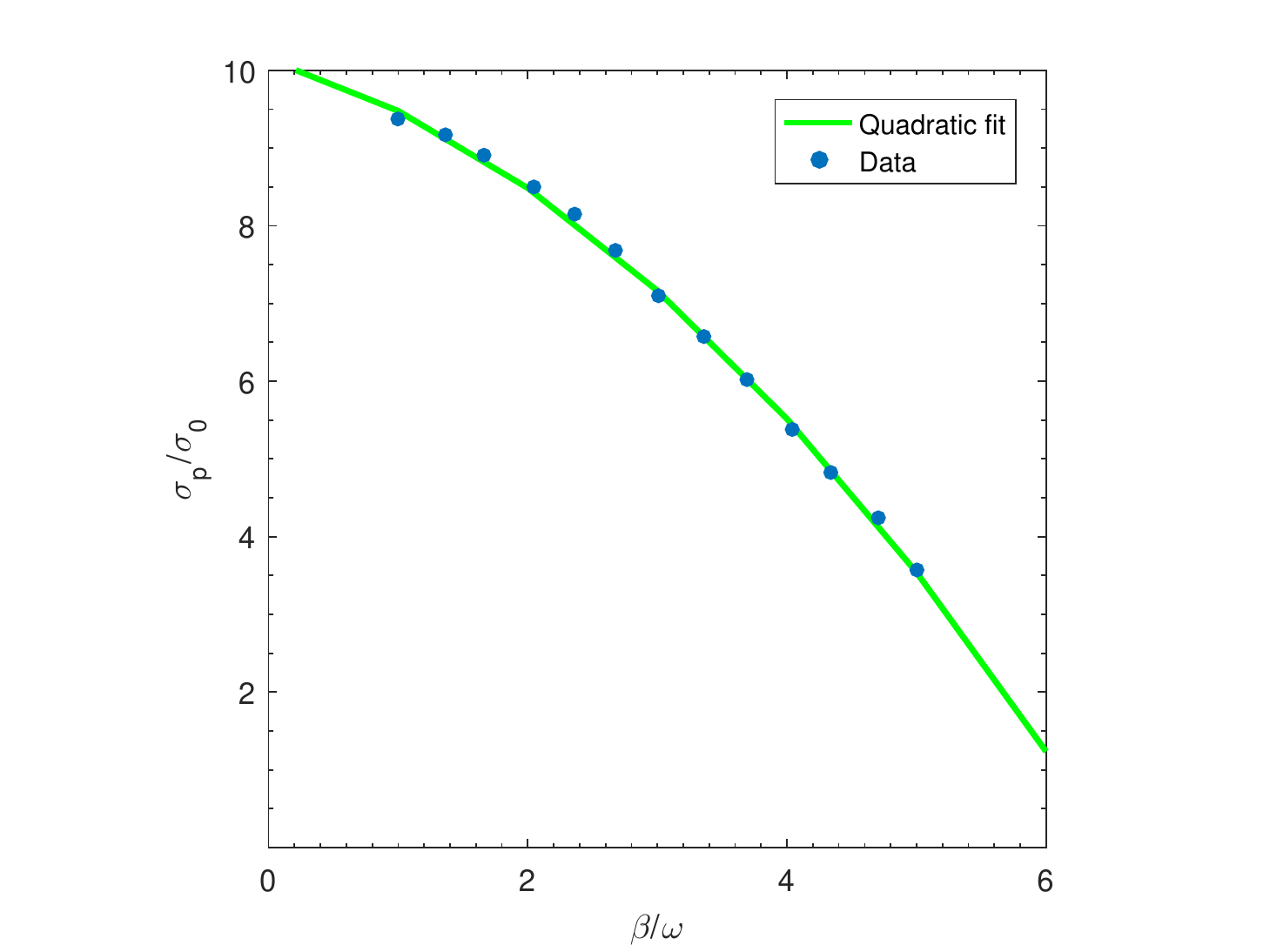}
\caption{\label{fig:9}Variance in the pressure measured by the simulated barometer in terms of $\beta/\omega$. A quadratic polynomial is fitted to the data with $R^2\approx 0.999$ }

\end{figure}

We remark that in a gas system such as an ideal gas, if the particles reflect perfectly from the container's walls, then the system will not travel to different parts of phase space. For example in the case of an ideal gas, the magnitude of the momentum of each particle remains unchanged. Hence the container plays an important role in perturbing the system and connecting these isolated parts of phase space together.

As we described in section II, we have assumed all of the collisions are elastic and therefore the inelastic collisions will not glue the isolated parts of phase space together anymore. Then one may ask what kind of perturbation pushes the system towards equilibrium. In our case, the main amount of perturbation is due to the computational errors. So the time scale in which system travels to the different parts of phase space, or equivalently the time scale in which thermal fluctuations arises, would be far bigger than the time scale of collisions in which we measure the pressure. Hence we can safely conclude that this fluctuation is mostly related to the instrumental part rather than thermal fluctuation.

\section{Thermal Fluctuations}
In this section we will derive a theoretical expression for the thermal fluctuation, based on the state equation \ref{stateeq}. According to \cite{4}, the probability of finding the system somewhere near its equilibrium point due to thermal fluctuation, $Pr$, is given as,
\begin{align}
Pr&=Pr_0 e^{-(\Delta G_0/kT) }\nonumber\\
G_0&=U+p_0 V-T_0 S\label{thermal}
\end{align}

where $p_0$ and $T_0$ are respectively the equilibrium pressure and temperature and $G_0$ is called the availability of the system. By expanding $G_0$ up to the second order in terms of $\Delta p$ and $\Delta T$ we have,
\begin{align}
\Delta G_0&=(\frac{\partial G_0}{\partial p})_T \Delta p+(\frac{\partial G_0}{\partial T})_p \Delta T\nonumber\\
&+(\frac{1}{2} \frac{\partial^2 G_0}{\partial p^2})_T \Delta p^2+\frac{1}{2} (\frac{\partial^2 G_0}{\partial T^2})_p \Delta T^2+(\frac{\partial^2 G_0}{\partial T\partial p}) \Delta p \Delta T\label{taylorexp}
\end{align}
Note that all of our calculations are taking place around the equilibrium point where the first derivatives of $G_0$ vanish. By substituting zero for these first derivatives in equation \ref{taylorexp} we have,
\begin{align}
\Delta G_0=(\frac{1}{2} \frac{\partial^2 G_0}{\partial p^2})_T \Delta p^2+\frac{1}{2} (\frac{\partial^2 G_0}{\partial T^2})_p \Delta T^2+(\frac{\partial^2 G_0}{\partial T\partial p}) \Delta p \Delta T\label{taylorexp2}
\end{align}
The next step is to find $G_0$ in terms of $p$ and $T$. For this we need to find entropy in terms of pressure and temperature. According to the first law of thermodynamics we have:
\begin{equation}
dU=TdS-pdV
\end{equation}
Using the state equation \ref{stateeq} and replacing $U$ in terms of $p$ and $T$ we have,
\begin{equation}
ds=\frac{dp}{kp}+\frac{(a+1)dV}{k(aV-b)}\label{ds}
\end{equation}
By integrating both sides of the equation \ref{ds} and replacing $V$ in terms of $p$ and $T$ from equation \ref{stateeq}, we have,
\begin{equation}
\Delta S=-\frac{1}{ka} \Delta ln⁡(p)+\frac{1+1/a}{k} \Delta ln⁡(T)
\end{equation}
Now we have both $S$ and $V$ in terms of $p$ and $T$. By substituting $S$ and $V$ in terms of $p$ and $T$ one can find $G_0$ in terms of $p$ and $T$ as follows,
\begin{equation}
G_0=(\frac{T}{k}-\frac{(T_0 (1+1/a))}{k} ln⁡(T) )+\frac{T_0}{ka} ln⁡(p)+\frac{T p_0}{akp}+\frac{p_0 b}{a}
\end{equation}
Using this explicit formula for the availability, its second derivatives could be obtained as follows,
\begin{align}
(\frac{\partial^2 G_0}{\partial p^2 })_T&=\frac{2p_0T}{akp^3} -\frac{T_0}{akp^2} \nonumber\\
(\frac{\partial^2 G_0}{\partial T^2 })_p&=\frac{T_0 (1+1/a)}{kT^2} \nonumber\\
(\frac{\partial^2 G_0}{\partial T\partial p})&=-\frac{p_0}{akp^2} \nonumber
\end{align}
Using the equation \ref{thermal} and by substituting second derivatives of $G_0$ in the equation \ref{taylorexp2}, one can obtain the probability distribution ($Pr$) as follows,
\begin{align}
Pr=Pr_0 e^{-x\Delta p^2-y\Delta T^2+z\Delta p\Delta T}
\end{align}
where,
\begin{align}
x=\frac{1}{2ak^2 p_0^2}~,~y=\frac{1+1/a}{k^2T_0^2}~,~z=\frac{1}{ak^2 p_0T_0}\nonumber
\end{align}
This function is not Gaussian and one might worry about its normalizability. For example, in figure 10 you can see the diagram of $f(x,y)=exp(-(x-5)^2-(y-5)^2+3(x-5)(y-5))$ which is saddle like and is not normalizable.
\begin{figure}[H]
\centering
\includegraphics[width=0.6\linewidth]{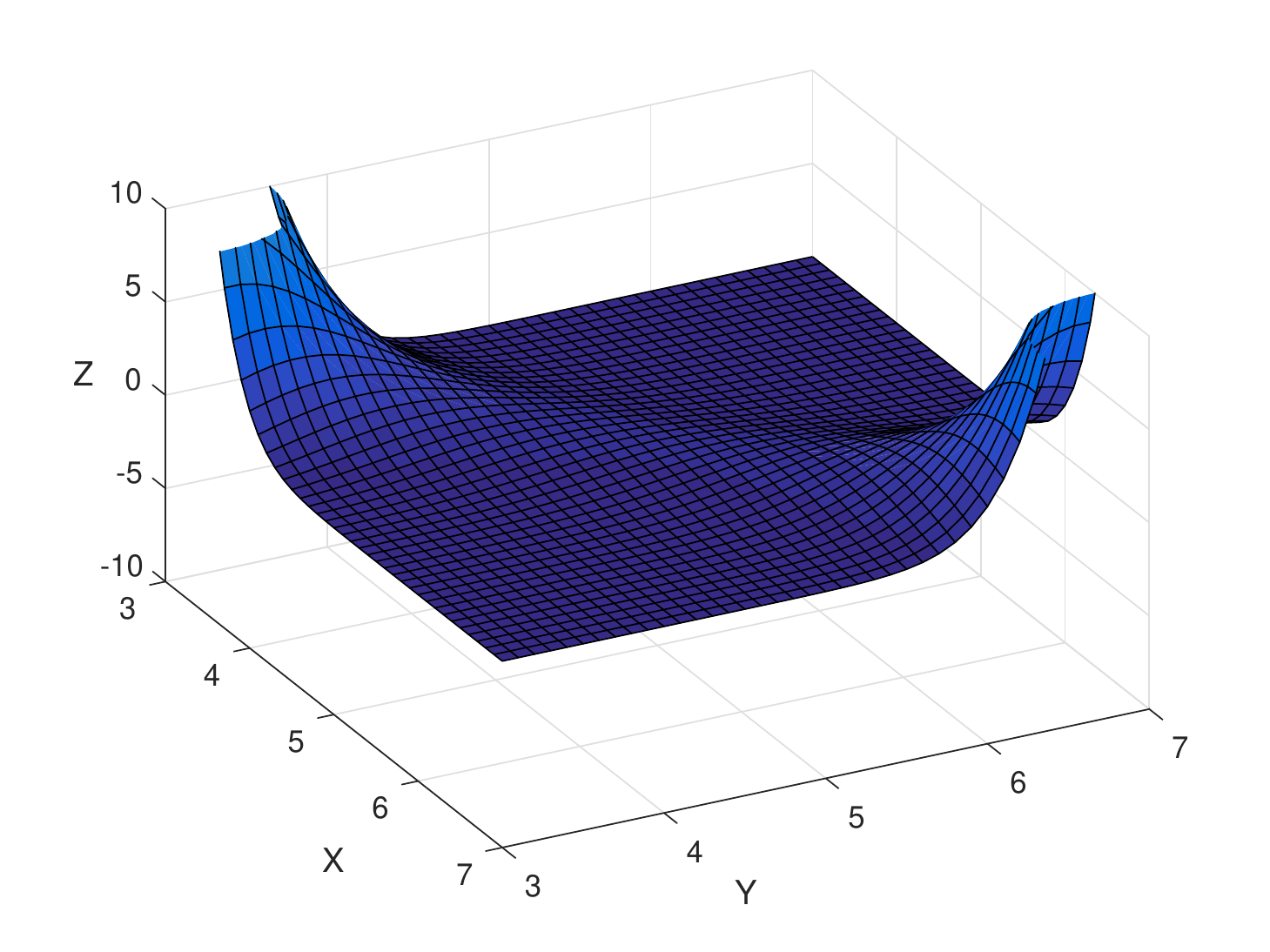}
\caption{\label{fig:10}The function $f(x,y)=e^{-(x-5)^2-(y-5)^2+3(x-5)(y-5)}$ }

\end{figure}

But in figure 11, we plot the diagram of a similar function with smaller $\Delta x\Delta y$ coefficient in the exponent which this time is normalizable (figure 11).

\begin{figure}[H]
\centering
\includegraphics[width=0.6\linewidth]{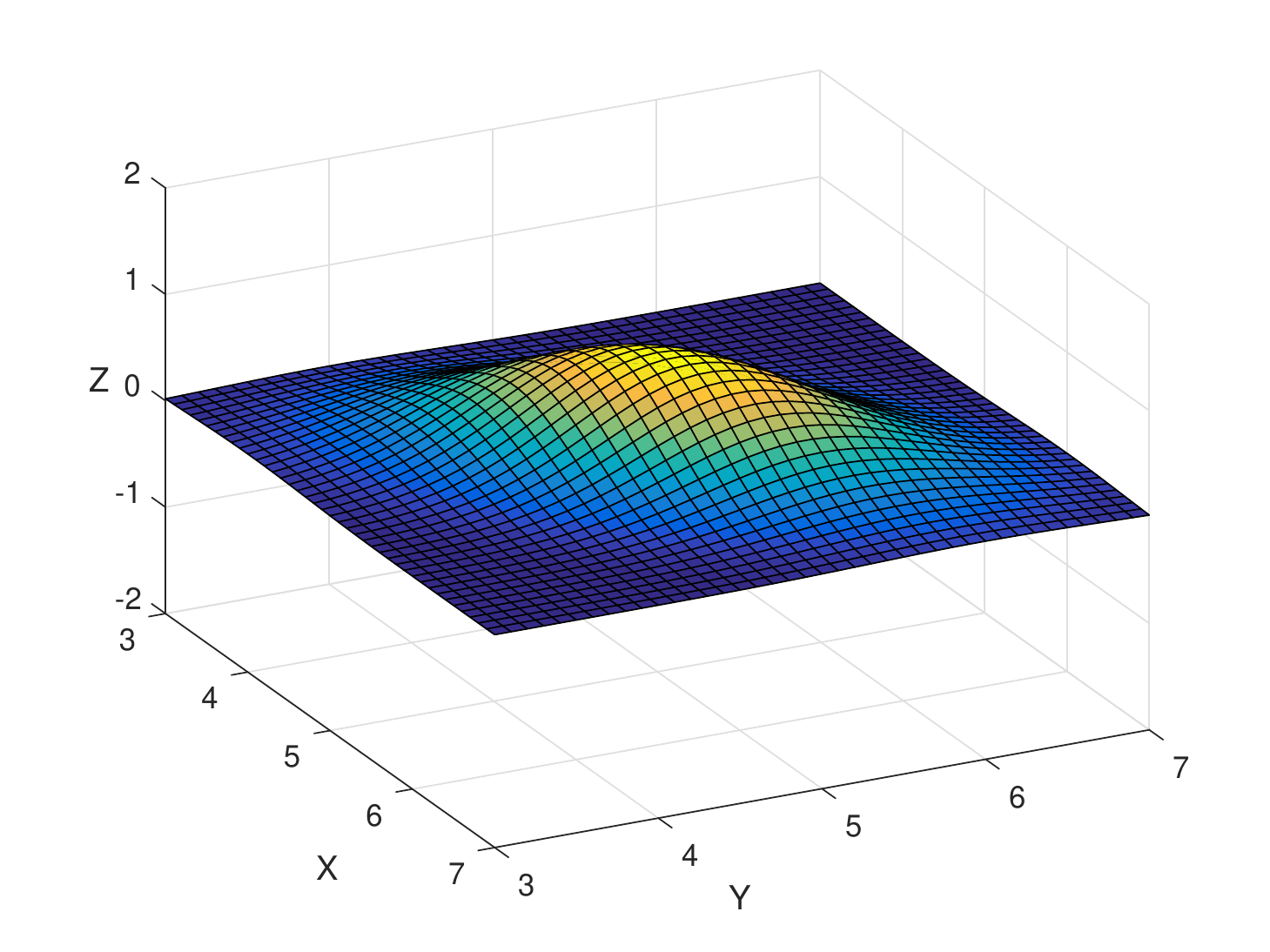}
\caption{\label{fig:12}The function $f(x,y)=e^{-(x-5)^2-(y-5)^2+(x-5)(y-5)}$ }

\end{figure}
In fact the sufficient and necessary condition for a function $f(x,y)=exp(-A\Delta x^2-B\Delta y^2+C\Delta x\Delta y)$ to be normalizable, is $2\sqrt{AB}>C$. Here we check this condition for $Pr⁡(p,T)$:
\begin{align}
2\sqrt{xy}>z&\leftrightarrow 2\sqrt{\frac{1}{2ak^2 p_0^2 }\times\frac{1+1/a}{k^2 T_0^2 }}>\frac{1}{ak^2 p_0 T_0 }\nonumber\\
&\leftrightarrow 2\sqrt{\frac{1+1/a}{2ak^4 p_0^2 T_0^2 }}>\frac{1}{ak^2 p_0 T_0 }\nonumber\\
&\leftrightarrow \frac{2(1+1/a)}{a}>\frac{1}{a^2} \leftrightarrow a>-\frac{1}{2}\nonumber
\end{align}
So $Pr(\Delta p,\Delta T)$ is normalizable for every positive value of $a$. By integrating the probability distribution, $Pr(\Delta p,\Delta T)$, over different values of $\Delta T$, we derive the probability of having pressure fluctuation $\Delta p$, despite the fluctuation of temperature as follows,
\begin{align}
Pr_{Th.}(\Delta p)=\int_{-\infty}^\infty Pr⁡(\Delta p,\Delta T)d\Delta T=me^{-n\Delta p^2 }\label{thermaldist}
\end{align}
where,
\begin{align}
m=Pr_0 \sqrt{\pi/y},n=x-z^2/4y\nonumber
\end{align}

\section{Overall Fluctuation in Experimental Data}

In this section, we will combine the thermal and instrumental fluctuations which we have obtained in the previous sections. We find the overall probability of finding $\Delta p$ deviation in measurement of pressure by $Pr_{Exp.}(\Delta p)$. According to the equations \ref{instrumentaldist} and \ref{thermaldist}, the probability distribution resulted by instrumental error ($Pr_{Inst.}$) and natural fluctuations ($Pr_{Th.}$) can be shown as,
\begin{align}
Pr_{Inst.} (\Delta p)=m' e^{-n' \Delta p^2 }\nonumber
\end{align}
where,
\begin{align}
n'=(c_2 (\omega) \beta^2+c_1 (\omega)\beta+c_0 (\omega))^{-2}\nonumber
\end{align}
and
\begin{align}
Pr_{Th.} (\Delta p)=me^{-n\Delta p^2 }\nonumber
\end{align}
where,
\begin{align}
n=x-z^2/4y;\label{flucts}
\end{align}
In the appendix \ref{app:3}, we have shown that how these two Gaussian probability distributions could be merged. The probability related to the overall fluctuation in measured data could be written as,
\begin{align}
&Pr_{Experiment} (\Delta p)=\rho e^{-\gamma \Delta p^2 };\nonumber\\
&\rho=mm'\sqrt{\frac{\pi}{n+n' }} ,\gamma=\frac{1}{\frac{1}{n}+\frac{1}{n'}}\label{total}
\end{align}
Using equations \ref{flucts} and \ref{total}, the overall variance of data which will be observed by an experimenter takes the following form,
\begin{align}
\sigma&=\sqrt{\frac{1}{2\gamma}}=\sqrt{\frac{1}{2n}+\frac{1}{2n'}}\nonumber\\
&=\sqrt{\frac{2k^2 p_0^2 a(a+1)}{(2a+1)}+\frac{(c_1 (\omega) \beta^2+c_1 (\omega)\beta)^2}{2}}
\end{align}
\section{Conclusions}
By introducing a novel simulation method and using classical thermodynamics, we obtained formulas for the thermal fluctuation and instrumental error separately. By combining them together, We derived a formula for the final variance which an experimenter will observe in the measured data. We remark that our results can be experimentally verified. Using this method one has the advantage of distinguishing and studying the thermal fluctuation which is of a physical importance.

\begin{acknowledgments}
We are very grateful to Olinka Bedroya and Hessamadin Arfaei for boosting this work by providing their useful advises and fruitful contributions. 
\end{acknowledgments}

\appendix
\section{}\label{app:1}
From the point of statistical mechanics, it is reasonable to assume that temperature, $T$, is linear in terms of pressure, $p$, as both have a linear dependence on the average kinetic energy \cite{5}. Also please note that we are considering a gas in which the electrostatic interactions lead to a significant increase in its pressure, such that we can neglect the intercept and assume that $T$ is proportional to $p$.
\begin{equation}
T=pf(V)\label{A1}
\end{equation}
Next we will obtain the relation between $p$ and $V$ along an adiabatic expansion. Note that $dW$ stands for the differential of work which has been done on the system. Since the process is adiabatic we have,
\begin{equation}
dQ=0\rightarrow dU=dW=-pdV\nonumber
\end{equation}
By substituting $U$ in terms of $p$ and $V$ using equation \ref{stateeq} we have,
\begin{align}
-pdV=dU&=(aV-b)dp+apdV \nonumber\\
&\rightarrow (a+1)dV/(aV-b)+dp/p=0\nonumber\\
&\rightarrow (aV-b)^{(1+1/a)} p=cte\label{A2}
\end{align}
In order to simplify the calculations, define:
\begin{equation}
g(p,V)=(aV-b)^{(1+1/a) } p\nonumber
\end{equation}
Now consider a Carnot cycle in which system goes under two adiabatic evolutions and two isothermal evolutions in the cycle. Take the first adiabatic evolution to be between states $(p_2,V_2 )$ and $(p_3,V_3)$ and the second evolution to be between $(p_4,V_4 )$ and $(p_1,V_1)$. Also consider that the first isothermal evolution is taking place between states $(p_1,V_1 )$ and $(p_2,V_2)$ with fixed temperature $ T_1$ and the second one between states $(p_3,V_3 )$ and $(p_4,V_4) $ with the fixed temperature $T_2$.

First and third evolutions in the cycle are isothermal, so according to \ref{A1} we have,
\begin{equation}
p_1 f(V_1 )=p_2 f(V_2 ),~~p_3 f(V_3 )=p_4 f(V_4)
\end{equation}
Second and forth evolutions are adiabatic so according to \ref{A2} we have,
\begin{equation}
g(p_2,V_2 )=g(p_3,V_3 ),~~g(p_4,V_4 )=g(p_1,V_1 )
\end{equation}
Rewriting the first law of thermodynamics for an isothermal expansion results:
\begin{align}
dU=dQ-pdV\rightarrow dQ=dU+pdV=dU+(\frac{T}{f(V)} )dV\label{A5}
\end{align}
By integrating equation \ref{A5} along an isothermal path we have,
\begin{align}
\Delta Q=\Delta U+\int_i^f pdV&=\Delta U+\int_i^f\frac{T}{f(V)} dV\nonumber\\
&=\Delta(U+F(V))\label{A6}
\end{align}
where,
\begin{align}
F=\int\frac{1}{f} dV\nonumber
\end{align}
Using equation \ref{A6} and the second law of thermodynamics one can write the following equation for the Carnot cycle,
\begin{align}
T_1/T_2 &=Q_{in}/Q_{out}\nonumber\\
 &=\frac{\Delta(U+F(V))|_1^2}{\Delta(U+F(V))|_3^4 }\nonumber\\
&=\frac{p_2 (aV_2-b)-p_1 (aV_1-b)+T_1 (F(V_1 )-F(V_2 ))}{p_4 (aV_4-b)-p_3 (aV_3-b)+T_2 (F(V_3 )-F(V_4 )) }\nonumber\\
&=\frac{R(p_1,V_1,p_2,V_2 )}{R(p_3,V_3,p_4,V_4 )}
\end{align}
where,
\begin{align}
R(x,y,z,t)=z(at-b)-x(ay-b)+xf(y)(F(y)-F(t))\label{A8}
\end{align}
Points $(p_1,V_1 )$ and $(p_2,V_2)$ of the Carnot cycle, can be determined uniquely by $T_1,g(p_1,V_1 )$ and $g(p_2,V_2)$ since they are the intersections of the isothermal path with constant temperature $T_1$ with the adiabatic paths which could be determined by their $g(p,V)$ value. Hence $Q_{in}$ is a function of $g(p_1,V_1 ), g(p_2,V_2)$ and $T_1$ . Furthermore, we know that for a Carnot cycle, $Q_{in}$ is proportional to $T_1$ so it can be written as,
\begin{equation}
Q_{in}=T_1 h(g(p_1,V_1 ),g(p_2,V_2 ))
\end{equation}
Hence if $(p_1,V_1 ),(p_2,V_2 )$ are located on an isothermal path, then we have,
\begin{align}
R(p_1,V_1,p_2,V_2 ):&=Q_{in}\nonumber\\
&=p_1 f(V_1 )h(g(p_1,V_1 ),g(p_2,V_2 ))\label{A10}
\end{align}
Multiplying both $p_1$ and $p_2$ by a constant $c$, will not change the equality $p_1 f(V_1 )=p_2 f(V_2 )$ so $(cp_1,V_1 ),(cp_2,V_2)$ will still lie on an isothermal path. On the other hand $R$ is a linear function in terms of $p_1$ and $p_2$ so we have,
\begin{equation}
R(cp_1,V_1,cp_2,V_2 )=cR(p_1,V_1,p_2,V_2 )\nonumber
\end{equation}
\begin{align}
\rightarrow cp_1 f(V_1 )&h(g(cp_1,V_1 ),g(cp_2,V_2 ))\nonumber\\
&=R(cp_1,V_1,cp_2,V_2 )\nonumber\\
&=cR(p_1,V_1,p_2,V_2 )\nonumber\\
&=cp_1 f(V_1 )h(g(p_1,V_1 ),g(p_2,V_2 ))\nonumber
\end{align}
\begin{equation}
\rightarrow h(g(p_1,V_1 ),g(p_2,V_2 ))=h(g(cp_1,V_1 ),g(cp_2,V_2 ))
\end{equation}
On the other hand the function $g$ is also linear in terms of $p$ so we have,
\begin{equation}
\rightarrow h(g(p_1,V_1 ),g(p_2,V_2 ))=h(cg(p_1,V_1 ),cg(p_2,V_2 ))
\end{equation}
This means $h(A,B)=h(cA,cB)$ for arbitrary positive values of A and B. Hence $h$ is a function of $x/y$.
\begin{equation}
h(x,y)=\alpha(\frac{x}{y})\label{A13}
\end{equation}
Rewriting the function $R$ using equations \ref{A8}, \ref{A10} and \ref{A13} we have,
\begin{align}
p_2 (aV_2-b)-p_1 (aV_1-b)&+T_1 (F(V_1 )-F(V_2 ))\nonumber\\
&=R(p_1,V_1,p_2,V_2 )\nonumber\\
&=cp_1 f(V_1 )h(g(p_1,V_1 ),g(p_2,V_2 ))\nonumber\\
&=pf(V)\alpha(\frac{g(p_1,V_1 )}{g(p_2,V_2 )} )\nonumber
\end{align}
\begin{equation}
\rightarrow \frac{aV_2-b}{f(V_2 )} -\frac{aV_1-b}{f(V_1 )} +(F(V_2 )-F(V_1 ))=\alpha(\frac{g(p_1,V_1 )}{g(p_2,V_2 )})\label{A14}
\end{equation}
Using the explicit expression of $g$ we have,
\begin{equation}
\frac{g(p_1,V_1 )}{g(p_2,V_2 )} =\frac{\theta(V_1 )}{\theta(V_2 )} \label{A15}
\end{equation}
where,
\begin{equation}
\theta(V)=\frac{(aV-b)^{(1+1/a)}}{f(V)}\nonumber
\end{equation}
By substituting equation \ref{A14} in \ref{A15} we have,
\begin{equation}
\Delta(\frac{(aV-b)}{f(V)} +F(V))=\xi(\Delta ln⁡(\theta(V))\label{A16}
\end{equation}
where $\xi(x)=-\alpha(e^x )$.

Therefore $\xi$ is an additive function and as a result it is linear. By substituting $\xi(x)=c_1 x$ in equation \ref{A16} we have,
\begin{equation}
\Delta(\frac{(aV-b)}{f(V)} +F(V))=c_1 (\Delta ln⁡(\theta(V)) )\label{A17}
\end{equation}
According to the equation (\ref{A17}) we have the following equation for a constant $c_2$.
\begin{equation}
\frac{(aV-b)}{f(V)} +F(V)=c_1 ln⁡(\theta(V))+c_2\label{A18}
\end{equation}
By substituting $\theta(V)$ in the equation \ref{A18} we have,
\begin{equation}
\frac{(aV-b)}{f(V)} +F(V)=c_1 ln⁡(\frac{(aV-b)^{(1+1/a)}}{f(V) })+c_2\label{A19}
\end{equation}
Replacing $\frac{d}{dV}F(V)$ by $\frac{1}{f(V)}$ based on the definition of $F(V)$, which is defined as $\int\frac{1}{f(V)}dV$, leads to,
\begin{equation}
(aV-b)\frac{d}{dV} F+F=c_1 ln⁡((aV-b)^{(1+1/a)} \frac{d}{dV}F )+c_2
\end{equation}
The solution of this ODE for $F(V)$, is a linear function in terms of $ln(aV-b)$ so one can find $f$ as following,
\begin{align}
&\frac{1}{f}=\frac{d}{dV}F=\frac{cte}{aV-b}\nonumber\\
&\rightarrow \frac{f(V)}{aV-b}=cte\nonumber\\
&\rightarrow \frac{pf(V)}{p(aV-b)}=cte\nonumber\\
&\rightarrow \frac{T}{U}=cte
\end{align}
which is the desired result.

\section{}\label{app:2}

It can be easily verified that $x=A (e^{(-\beta+\sqrt{\beta^2-\omega^2 })t}-e^{(-\beta-\sqrt{\beta^2-\omega^2 })t})$ is a solution to the equation \ref{oscillator}. We have to find the appropriate constant $A$ which normalizes this function.
\begin{align}
&\int_0^\infty A (e^{(-\beta+\sqrt{\beta^2-\omega^2 })t}-e^{(-\beta-\sqrt{\beta^2-\omega^2 })t})dt=1\nonumber\\
&\rightarrow A(\frac{1}{-\beta+\sqrt{\beta^2-\omega^2 }} +\frac{1}{-\beta-\sqrt{\beta^2-\omega^2 } })\nonumber\\
&\rightarrow A(\frac{2 \sqrt{\beta^2-\omega^2 }}{\omega^2} )=1\nonumber\\
&\rightarrow A=\frac{\omega^2}{2\sqrt{\beta^2-\omega^2 }}\nonumber\\
\end{align}
So the normalized solution of equation \ref{oscillator} is given as,
\begin{align}
x(t)=\frac{\omega^2}{2\sqrt{\beta^2-\omega^2 }}(e^{(-\beta+\sqrt{\beta^2-\omega^2 })t}-e^{(-\beta-\sqrt{\beta^2-\omega^2 })t})
\end{align}

\section{}\label{app:3}

In order to find the probability of reporting $p+\Delta p$ as the measured pressure, we have to consider any situation in which an amount of $x$ from the total deviation $\Delta p$ is due to the thermal fluctuation and the remaining is because of the instrumental error. Hence we have to multiply the probabilities of the first event ($Pr_{Th.}(⁡x)$ ) and the second event ($Pr_{Inst.}(⁡\Delta p-x)$ ) and integrate it over $x$.
\begin{align}
&Pr_{Experiment} (\Delta p)\nonumber\\
&=\int_{-\infty}^\infty me^{-nx^2 } m' e^{-n' (\Delta p-x)^2 } dx\nonumber\\
&=\int_{-\infty}^\infty mm'(e^{-nx^2-n' (\Delta p-x)^2 } ) dx\nonumber\\
&=\int_{-\infty}^\infty me^{-(n+n' ) x^2-n' \Delta p^2+n' x\Delta p} dx\nonumber\\
&=\int_{-\infty}^\infty me^{-(n+n' ) (x-\Delta p n'/(n+n' ))^2+(nn')/(n+n' ) \Delta p^2 } dx\nonumber\\
&=mm'\sqrt{\frac{\pi}{n+n' }} e^{-nn'/(n+n') \Delta p^2 }
\end{align}
which is the desired result.





\bibliography{ref}
\end{document}